\documentclass{article}
\usepackage[margin=1.0in]{geometry}
\usepackage{subfigure,epsfig,amsfonts}
\usepackage{natbib}
\usepackage{amsmath}
\usepackage{amssymb}
\usepackage{amsthm}

\usepackage{listings} 
\usepackage{enumitem} 
\usepackage{lineno, blindtext} 
\usepackage{xstring} 
\usepackage{longtable}
\usepackage{color}
\usepackage{multicol}
\usepackage{url}

\usepackage{mathtools}

\newcommand{\fontit}[1]{\textit{#1}}

\newcommand{\name}{EIN}

\newcommand{\secref}[1]{Section~\ref{#1}}

\newcommand{\figref}[1]{Figure~\ref{#1}}

\definecolor{Black}{rgb}{0.0,0.0,0.0}
\definecolor{ngray}{rgb}{0.5,0.5,0.5}

\lstdefinelanguage{Diderot}{%
  morekeywords={%
    bool,%
    die,%
    else,%
    false,field,foreach,%
identity,if,image,load,in,initially,input,int,%
    fem,
    kernel,%
    nan,new,%
    output,%
    real,%
    function,stabilize,strand,string,%
    inside,tensor,true,%
    update,%
    vec2,vec3,vec4,%
    zeros},
  sensitive,%
  morecomment=[s]{/*}{*/},%
  morecomment=[l]//,
  morestring=[b]"}%

\title{An exploration to visualize finite element data with a DSL}
\author{Charisee Chiw, Gordon Kindlmann, and John Reppy}

\begin{document}
 \maketitle
 \abstract{
 The scientific community use PDEs to model a range of problems. 
The people in this domain are interested in visualizing their results, but existing mechanisms for visualization can not handle the full richness of computations in the domain.
We did an exploration to see how Diderot, a domain specific language for scientific visualization and image analysis [8,13], could be used to solve this problem.

We demonstrate our first and modest approach of visualizing FE data with Diderot and  provide examples.
Using Diderot, we do a simple sampling and a volume rendering of a FE field.
These examples showcase Diderot's ability to provide a visualization result for Firedrake. 
This paper describes the extension of the Diderot language to include FE data. 
}

\section{Introduction}


Computational scientists compute solutions to systems of partial differential equations (PDEs) on large finite meshes using numerical techniques, such as the finite element method (FEM). 
These PDEs can be used to describe complex phenomena like turbulent fluid flow.
The solution to PDEs or the output to software that solves PDEs are sometimes referred to as ``FE fields".
The Diderot language does not know how to solve PDEs, or represent FE fields, but it does have the computational model to visualize fields.
The work in this paper takes a step towards using Diderot to visualize FE fields.
With this work we hope to use Diderot to help debug visualizations of FE fields and enable more interesting visualizations.

Solving PDEs and visualizing PDEs require two different techniques and entirely different code.
To simplify the transition from solving a PDE to visualizing its solution,
scientists may turn to standard visualization tools to analyze their data. 
The problem with this approach is that there is not a universal solution to accurately visualize every PDE.
For example, visualizing finite-element data created with higher order elements and a small number of cells can lead to images that do not accurately represent the original solution.
We believe that Diderot can be useful.

We want the user to be able to use visualization programs enabled by Diderot on fields created by FEM.
Expecting users to transition from visualization toolkits to writing in a new programming languages and developing an expertise in scientific visualization is a big ask.
Our goal is to be able to augment any existing Diderot program (written for discrete data) and apply it to FE data with minimal changes to the program.
That way Diderot could compile programs to extract interesting visualization features from FE data [13].

Our work demonstrates a modest step towards visualizing  FE data with Diderot.
This paper is organized as follows.
\secref{fem:back} offers some background about this problem area and  provides  a motivating example.
 \secref{fem:communication} describes the implementation details to our approach.
\secref{demonstration} demonstrates an application of our approach by providing one example of the Helmholtz equation and interpolating a function.
We end with a discussion in \secref{fem:discuss}.

\section{Motivation}
\label{fem:back}

The Finite Element Method [4] 
 gives a general framework for computing solutions to differential equations.
In \secref{fem:motivate:back}, we more closely describe FE fields and the software used to create them.
In \secref{fem:motivate} we describe the state of using visualization toolkits to visualize a certain domain of FE fields. 
We provide an example of a problem that can not be visualized with the current state of visualization.

\subsection{Background}
\label{fem:motivate:back}

\paragraph{Diderot}
Diderot is a language designed for scientific visualization and image analysis. 
Algorithms in this domain depend on a high level of math.
Diderot eases the transition from visualization algorithms to code by providing the user with a rich mathematical notation.
We recently provided examples of more mature visualizations [13] 
 that can created with Diderot due to developments with our intermediate representation \name{} [5,6].
The computational core to visualization programs involve tensor and field operators applied to fields. 
In Diderot, we have previously defined fields as the convolution between discrete image data and a kernel.
Our goal is to extend our definition of a field to include other types of data.

Diderot has approached evaluating the correctness in the Diderot compiler in two ways; an automated testing infrastructure [7] 
 and presenting properties of the rewriting system  [9].
There has yet to be an evaluation of an image, created by an outside source. 
In this paper we evaluate the results based on visualization principles [14].
%
%
%

\paragraph{FE fields}
FE fields  are created from a solving a PDE on a finite element mesh, which involves discretizing the domain into small finite mesh elements and using a set of basis functions (derived from the mesh) to span the domain space. 
These fields  approximate numeric solutions to PDEs.

The Python code builds on the description of the problem.
 There are a variety of different meshes that are built-in or could be created by outside tools.
Often, we use a unit square mesh.
\begin{lstlisting}[mathescape=true] 
    m = UnitSquareMesh(2, 2)
\end{lstlisting}
The code creates a 2x2 mesh of a square. Each smaller square is divided into two triangles for a total of eight elements. 
  The mesh is one of the arguments when defining a function space.
\begin{lstlisting}[mathescape=true] 
    V = FunctionSpace(m, $``P"$, K)
\end{lstlisting}
$``$P" refers to a family of finite element spaces (other families include $``$DP", $``$RT", and $``$BDM").
 The basis functions are linear when K=1 and cubic when K=3.

A FE field can be created from solving a PDE, but it can also be generated by interpolating an analytically defined expression. 
\begin{lstlisting}[mathescape=true] 
    f = Function(V).interpolate(Expression(exp))
\end{lstlisting}    
In this chapter, when we wish to create simple examples we choose to generate fields from expressions.

\paragraph{Software}
Computer scientists build software to solve PDEs that represent a wide range of problems.
On the surface the software (or programming language) ideally represents a high-level math notation that is easy to understand but under the hood it is more complicated.
The translation between the notation and computer code involves several steps and pieces of software. 
Solving a PDE involves  the discretization of differential equations and uses the finite element method to provide an approximate solution.
Optimizing the translation from PDE equation to approximate solution is pursed by many groups  
 including the \fontit{FEniCS  Project} [10] 
 and \fontit{Firedrake}  [19].

The FEniCS [10,17] 
 project is an automated system to find solutions for partial differential equations using the finite element method.
It enables users to employ a wide range of discretization  to a variety of PDEs.
On the surface it uses the Unified Form Language (UFL) 
[2],
 a domain-specific language to represent weak formulations of partial differential equations.
 UFL supports tensor algebra, high-level expressions with domain-driven abstraction. 
UFL does not provide the problem solving environment, instead it creates an abstract representation that is used by form compilers, such as the FEniCS Form compiler (\fontit{FFC}) [3],
 to  generate low-level code.
FFC can implement  tensor reduction for finite element assembly [16] 
and aims to accept input from any multilinear variational form and any finite element to generate efficient code.

Firedrake [19] 
 is a similar system that is also used to solve PDEs. 
In addition to UFL, it uses a modified version of FFC [3], 
\fontit{FIAT} [15], 
a \fontit{PyOP2} interface [20], 
 and  \fontit{COFFEE} [18] 
 \footnote{We describe the Firedrake project at the time of our work but Firedrake has developed new parts of their infrastructure:  \fontit{FINAT} augments FIAT and their new form compiler \fontit{TSFC} replaces FFC.}.
FFC is the FEniCS form compiler for generation of low-level C kernels from UFL forms.
FIAT is the finite element automatic tabulator. 
It presents an abstract description of elements and has a wide range of finite element families.
PyOP2 provides a framework for carrying out parallel computations on unstructured meshes. 
The COFFEE compiler optimizes the abstract syntax trees generated by FFC.

\subsection{Creating and Visualizing FE data} 
\label{fem:motivate}
There are a number of approaches to supporting scientific visualization. 
A common way is to use a toolkit such as the Visualization Toolkit (VTK) [21],
ParaView [1], 
and the Insight Toolkit (ITK) [12] 
In this domain it is important for the visualization tool to understand how the data is represented and that restraint limits the options available to the FEM user.
Toolkits are commonly used to visualize the solutions created by FEM software.

Existing practices to visualize FEM are insufficient.
The strategy to solve PDEs is very different from the algorithms used to  visualize the result.
 Firedrake uses a VTK file format for its visualization output.
 The format only supports linear and quadratic data.
  Firedrake takes the output and writes the output to a linear file format\footnote{ The images in this paper are created when Firedrake did an L$^2$ projection, but now Firedrake uses interpolation to generate linear output.}. 
   Paraview  might then accurately assume linear basis functions to represent the Firedrake output even though the original solution was created with higher-order elements.
  As a result  the image may not accurately  represent the PDE solution.

In this section, we present the problem with the current state of visualizing FE data by providing a simple example of a how a bug can occur.
We create a field using higher-order data with a small number of cells.
We present the results of using Paraview and Diderot side by side.
 The implementation process that enables the communication between Diderot and Firedrake is introduced in more detail in \secref{fem:communication}. 
\begin{figure}
\begin{lstlisting}[mathescape=true] 
    exp = $``$x[0] *x[0] *(1-x[0] )$"$
    m = UnitSquareMesh(2, 2)
    V = FunctionSpace(m, $``$P$"$, K)
    f = Function(V).interpolate(Expression(exp))
\end{lstlisting}
\caption[Define a field in UFL] {The above code is written in Python and used to define a field by interpolating an analytically defined  expression  given the function space.
We define a polynomial expression $x^2(1-x)$
and a unit square mesh (\lstinline!m!).
The function space \lstinline!V! is defined by a mesh (\lstinline!m!), the family of finite element spaces (\lstinline!P!), and the order of the polynomial (\lstinline!K!).
The field (\lstinline!f!)  is defined by interpolating an analytically defined  expression  given the function space  (\lstinline!V!).
The expression (\lstinline!exp!) is composed with linear basis functions (when \lstinline!K!=1) and a cubic basis function (when \lstinline!K!=3)  using a unit square mesh (\lstinline!m!). 
}
\label{fem:code1}
\end{figure}
\begin{figure}[] \begin{center} \includegraphics[height=2in] {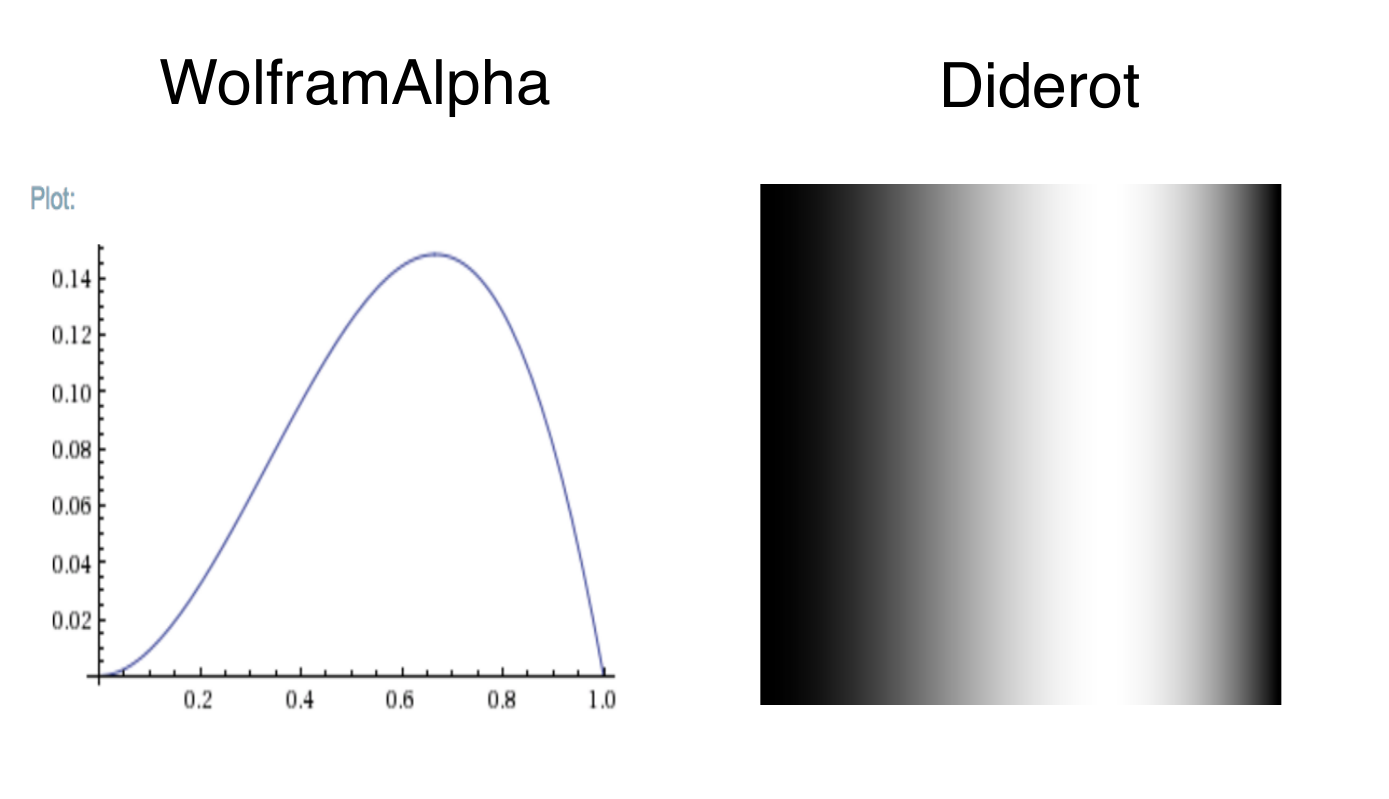}     
\caption[Expected results for creating and visualizing a field] {Fields created and visualized by a single source.  We defined a function $F(x) = x^2(1-x)$ where $x \in [0,1] $. We expect the maximal point to be at x= $\frac{2}{3}$.  WolframAlpha can quickly and easily graph the results.  In Diderot we synthesized a field by taking samples of a function defined by $F$ and saved the result to $out.nrrd$.  Then a second Diderot program was used to sample $out.nrrd$ and visualize the result.   }  \label{fig:fem:boxtruth}
\end{center}\end{figure}
\begin{figure}[] \begin{center}\includegraphics[height=3in] {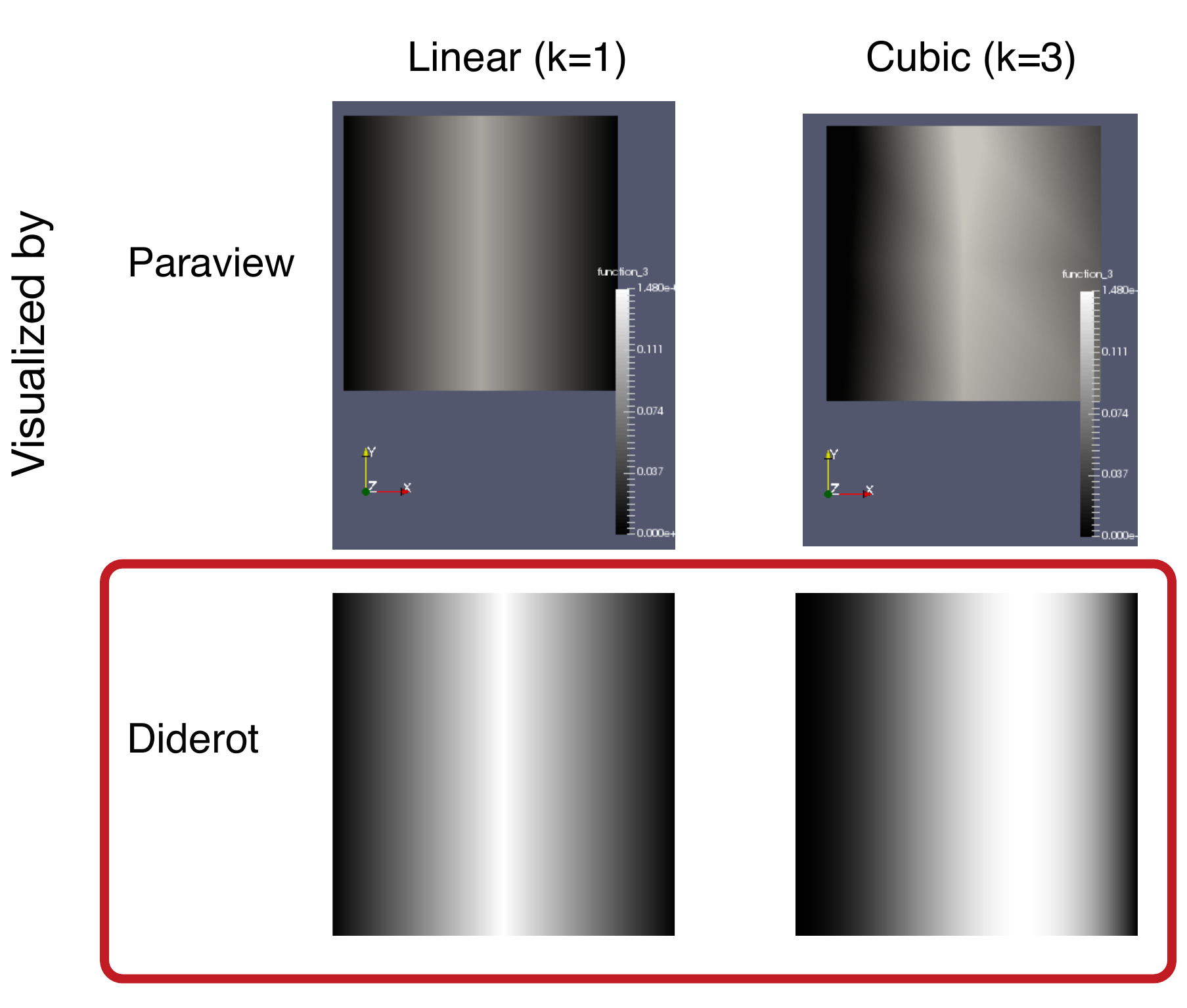}     
\caption[Visualizing FE data with different visualization tools] {Fields created by Firedrake using linear or cubic elements. We defined a function $F(x) = x^2(1-x)$ where $x \in [0,1] $. We expect the maximal point to be at x= $\frac{2}{3}$. In the following grayscale images, the maximal points are indicated by the brighter spots. The two left images use Firedrake data created with linear elements (K=1). Regardless of the visualization strategy we expect the images to be an inaccurate representation of the solution. The two images on the right   use Firedrake data created with cubic elements (K=3). The image (top right) created with Paraview is incorrect, and the image (bottom right) created with Diderot is correct.} \label{fig:fem:boxelse}
\end{center}\end{figure}
 \paragraph{Creating FE data}  
In the following example, we use Firedrake to create FE data.
Instead of solving a PDE, we  define a field  by interpolating the expression  given the function space.
We show the Python code in \figref{fem:code1}.

We define a polynomial expression $x^2(1-x)$.
Given a higher-order polynomial expression we can also assume that linear functions will not correctly be able to represent it. 
On the other hand, cubic functions should be able to offer a reasonable approximation for this problem. 

\paragraph{Visualizing FE data}  
We chose to visualize the expression several different ways in order to provide a means of comparison. \figref{fig:fem:boxtruth} and \figref{fig:fem:boxelse} present these visualizations.
The images in \figref{fig:fem:boxtruth} offer the ground truth for this example. 
\figref{fig:fem:boxelse} uses Firedrake data created with linear and cubic elements\footnote{As a note subdividing the field would create a more accurate solution but the context of the problem required creating a field with higher-order elements and a \fontit{small} number of cells.}. 
 Regardless of the visualization strategy we expect the images (left) created with linear elements to be an inaccurate representation of the solution.
 On the other hand, we expect the images created with cubic elements to be correct.
The image (bottom right) created with Diderot is correct  while the image (top right) created with Paraview is not.

When the results do not represent the field  it can be difficult to understand and use visualizations to debug.
From the user's perspective the issue could be with the user's UFL code, Firedrake's evaluation, or the visualization program. 
Our goal is to provide Diderot as an alternative tool that can be used in these instances.

 \section{Our Approach}\label{fem:communication}
It is not a long term solution but we created our current prototype to establish communication between Firedrake and Diderot. 
Firedrake creates FE data by solving a PDE or interpolating an expression over a function space.
Diderot visualizes the results and provides a few programs that can be initialized.
Firedrake provides a way to evaluate a field at a position and provides Diderot with a call back.
In the Diderot program a field is defined for FE data instead of using field reconstruction and convolution.
Everything else in the Diderot program is the same.
Lastly, when possible we try to evaluate the visualization based on established visualization principles.
We worked in collaboration with the Firedrake  team [19] 
 at the Imperial College of London. 
 \paragraph{FE data}
 A field is created using FE data with a Firedrake program.
 The field can be the result of solving a PDE or interpolating an expression over a function space.
 The Diderot syntax offers a limited way to represent tensor fields created by FE data.

 \paragraph{Diderot program}
 The core of visualization programs is independent of the source of data, but Diderot could only represent regular imaging grids. 
 We addressed this limitation by allowing the Diderot compiler to declare a field that is defined by FEM.
 Instead of creating a field by using field reconstruction and convolution 
the user can use the following Diderot line to indicate the new kind of field data.
\begin{lstlisting}[mathescape=true] 
    input fem#0(2)[] g;
    field#0(2)[] F = toField(g) 
\end{lstlisting} 
As a result, a visualization program written for regular imaging grids could be easily changed so it is applied to the result of a PDE solution.
We have created a few Diderot programs that have been compiled to a library.

 \paragraph{Firedrake program}
  The Diderot program provides the framework to sample fields and do volume rendering of 3-D fields. 
A Firedrake program makes a call to the Diderot library. 
A field reference is used to initialize a call to the Diderot library.
The following is a sample of Python code that would be written in a Firedrake program.
\begin{lstlisting}[mathescape=true] 
    res = 200 # resolution to MIP program
    step = 0.01 # step size to ray tracer
    diderot.mip(file_name, f, res, res, step)
\end{lstlisting}
It initializes visualization parameters (such as resolution and step size) and provides a pointer to the field.  
      
 \paragraph{Point Evaluation}
  Firedrake can evaluate fields (represented as a FE field) at arbitrary physical points [11] 
   It determines which cell to look at.
 Diderot does not know how to evaluate a FE field at a given position.
For each inside test, probe, and gradient operation the field has to be evaluated at a position.
The generated code  will use Firedrake's point evaluation functions for a given field and position.

\section{Demonstration}
\label{demonstration}

We provide two examples to demonstrate results of this work. 
In the first example, a field is created by interpolating an expression over an indicated function space and visualized with a volume rendering program. 
In the second example, we provide a classic example of a PDE and simply sample the result.

\subsection{Communication between Diderot and Firedrake}
\label{fem:comparediderot}

\begin{figure}[]  
 \begin{lstlisting}[mathescape=true] 
 vec3 camAt = [1,1,1] ;//position camera looks at
 input fem#0(3)[] g;
 field#0(3)[] F = toField(g);
 $\dots$
     if(!inSphere || |pos-camAt|< 1){out = max(out, F(pos));}
\end{lstlisting}
\caption[Augmented MIP program] {The above is Diderot code. Diderot is used to do a volume rendering of this 3-d field by setting up a camera and doing a MIP.
 Diderot allows a user to define a field by some external source. 
The Diderot program will generate code that will communicate to Firedrake's point evaluation capability to reduce \lstinline!F(pos)!.}
\label{fem:mip:code}
\end{figure}
\begin{figure}[]  
\includegraphics[width=5.2in] {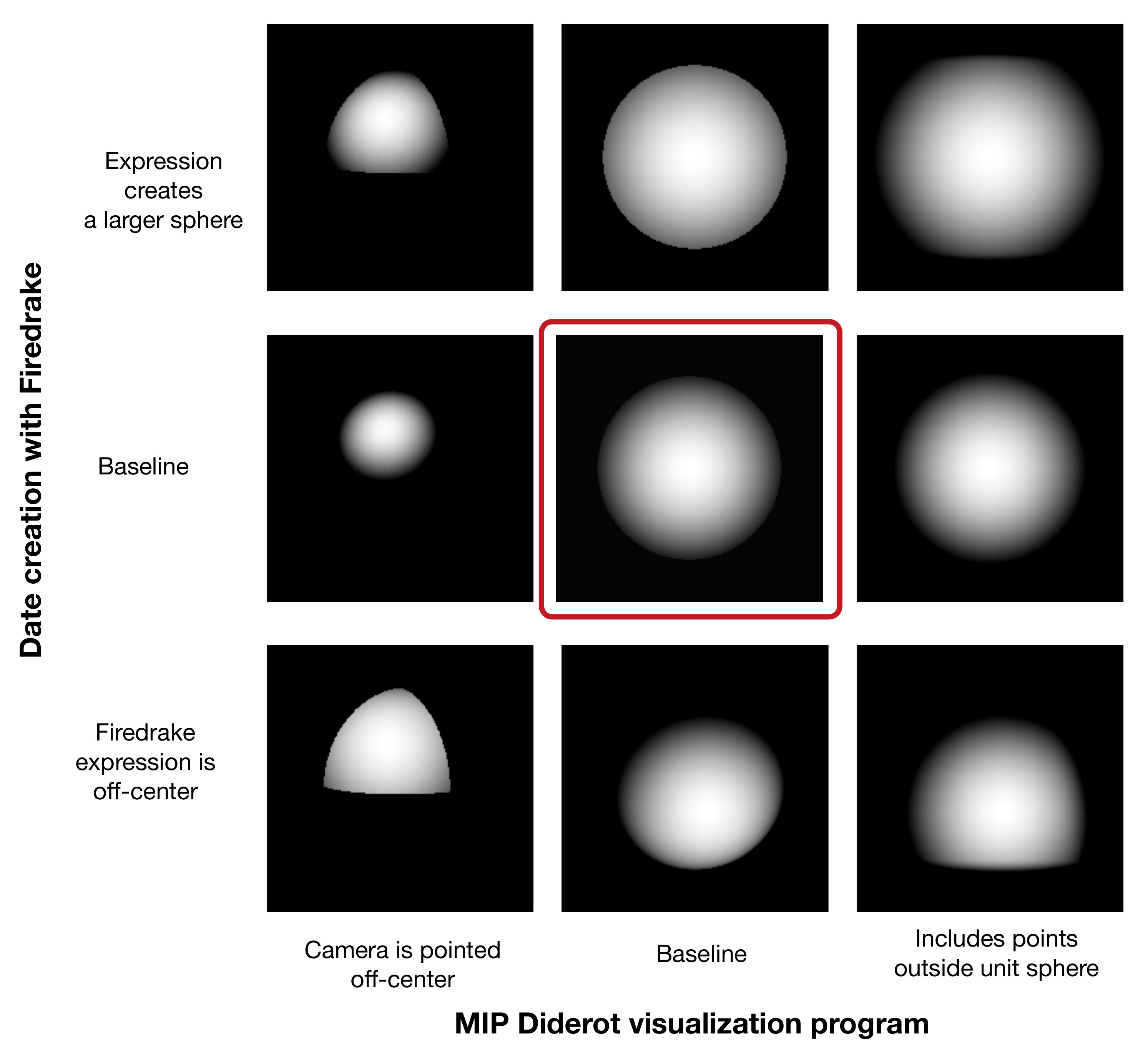}   
\caption[Establish communication between Diderot and Firedrake] {Creates sphere with Firedrake. Included are examples of bugs. The centered image is correct. }
  \label{fig:multiplefemsphere}
\end{figure}
\begin{figure}[] 
\includegraphics[width=6.3in] {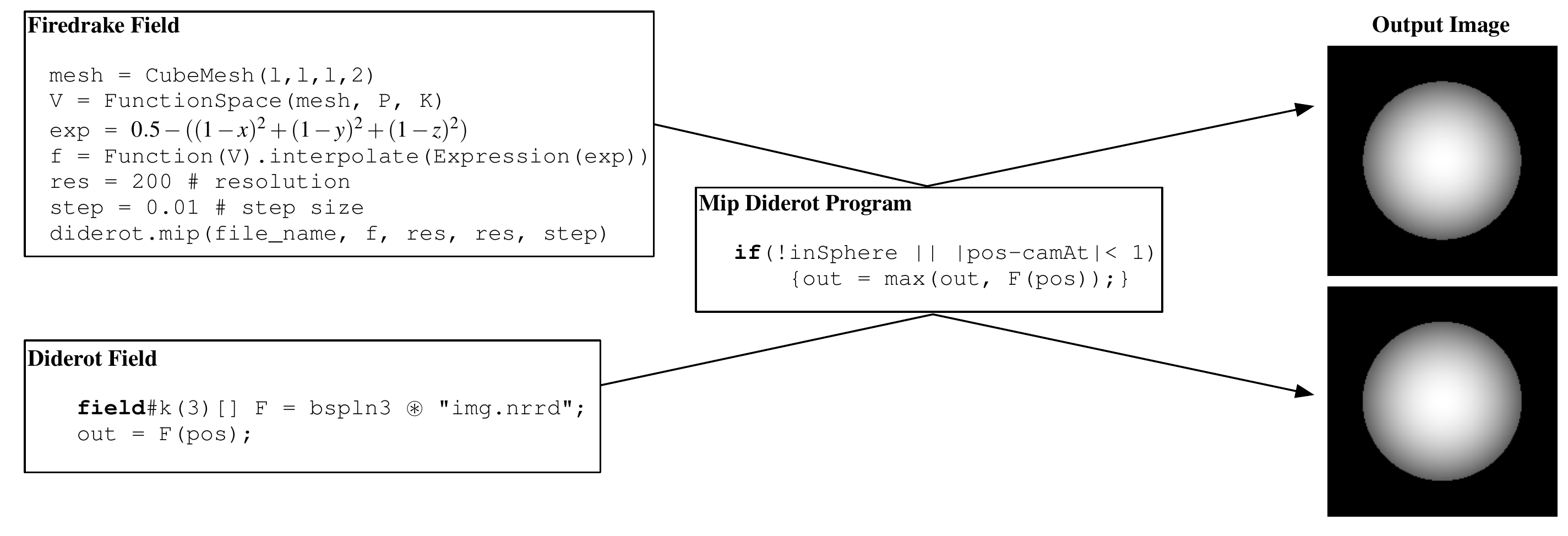}\\
\caption[Applying representation invariance principle] {This figure compares data created by two different sources and visualized with the same tool. The fields are created by Firedrake and Diderot. In the Firedrake code we create a cube mesh (\lstinline!m!) of length 2 on each side. 
The function space \lstinline!V! is defined by a mesh (\lstinline!m!), reference element (\lstinline!P!), and uses cubic polynomials.
The expression creates a symmetrical sphere centered at [1,1,1] and shifted by 0.5.
Field (\lstinline!f!) is defined by interpolating the expression given the function space  (\lstinline!V!) .
We make a call to Diderot library by passing the field and initializing the resolution and step size.
}
  \label{fig:fem:visver}
\end{figure}

\paragraph{Python code}
We create the field defined by interpolating the expression given the function space  using Firedrake. 
We show the Python code under the ``Firedrake Field" box in \figref{fig:fem:visver}.
We make a call to the Diderot library by passing the field and initializing the resolution and step size (for the Diderot program).

\paragraph{Diderot code}
Diderot visualizes the result by using an augmented MIP program.
MIP or maximum intensity projection is a minimal volume visualization tool for 3-d scalar images. 
The  Diderot code for the Diderot program is presented in \figref{fem:mip:code}.

\paragraph{Results}
We visualize the FE field using a Diderot implementation of a MIP program. 
We vary the
expression created by Firedrake, where the camera in the Diderot program is pointed, and how the data is included in the visualization program.
We try these different variations to mimic how a user might use an existing Diderot program (or template) to visualize their data.

We provide the results in \figref{fig:multiplefemsphere}.
We set the camera  to look at the center  of the data (center and right column) or off-center (left column).
We also created fields with the Firedrake expression centered (two top rows) and off-centered (third row).

The red box indicates the setting where the Diderot camera and firedrake expression are both centered and result is as expected (symmetric sphere).
The other images in the figure are subject to bugs, because the function space defined by Firedrake does not match the image space probed by Diderot (or where the camera is pointed).
This example illustrates the care that is needed when setting up the Firedrake Diderot pipeline.

\paragraph{Evaluate Visualization}
The data representation, either by FE field, or discrete field, should not change the visual representation (according to the Principle of Representation Invariance).
We compare the output for the field created by Firedrake and the field created by Diderot in \figref{fig:fem:visver}.   
As can be expected, the image created by Firedrake is essentially the same as that created by Diderot.

\subsection{PDE Example}
The Helmholtz problem is a symmetric problem and is a classic example of a PDE.
Consider the Helmholtz equation on a unit square $\Omega$ with boundary $\Gamma$.
$$-\nabla{^2} u + u =f$$
$$ \nabla u \cdot \vec{n} =0 \text{ on }\Gamma.$$
The solution to the equation is some function u $\in V$ for some suitable function space V that satisfies both equations.
After transforming the equation into weak form, applying a test function $V$, and integrating over the domain we get the following variational problem\footnote{The approach to solve PDEs with FEM is described in more detail in by Brenner and Scott [4].}.
\begin{equation} \int_{\Omega} \nabla u \cdot \nabla v + u v dx =\int_{\Omega} vf dx \label{helm:int}\end{equation}
 We choose function f\footnote{The example and code are provided by Firedrake} 
\begin{equation}
f=(1.0+8.0\pi^2)cos(2\pi x)cos(2\pi y)
\label{helm:f}
\end{equation}

\begin{figure}[]  
 \begin{lstlisting}[mathescape=true] 
    mesh = UnitSquareMesh(10, 10)
    V = FunctionSpace(mesh, $``$CG$"$, k)
    u = TrialFunction(V)
    v = TestFunction(V)
    f = Function(V)
    f = #$\text{Interpolate the expression }\eqref{helm:f}$     
    a = #$\text{Represents left-hand side of } \eqref{helm:int}$    
    L = #$\text{Represents right-hand side of }\eqref{helm:int}$    
    u = Function(V)
    solve(a == L, u, solver_parameters={'ksp_type': 'cg'})
    # $\text{Paraview output}$
    File($``$helmholtz.pvd$"$) << u
    # $\text{Call to Diderot}$
    res=100
    stepSize=0.01
    type=1  # creates nrrd file
    vis_diderot.basic_d2s_sample(namenrrd,u, res, stepSize, type)
\end{lstlisting}
\caption[Python code for Helmholtz problem] {We present the Python code to solve the Helmholtz problem. We omit some details and instead provide comments for clarity.}
\label{fem:helm:fire}
\end{figure}

\noindent 
The Python code (shown in \figref{fem:helm:fire}) solves the PDE and connects the solution to Diderot.
The original code uses linear elements (k=1), but we choose to use linear and cubic elements.

\begin{figure}[] 
\includegraphics[width=5.in] {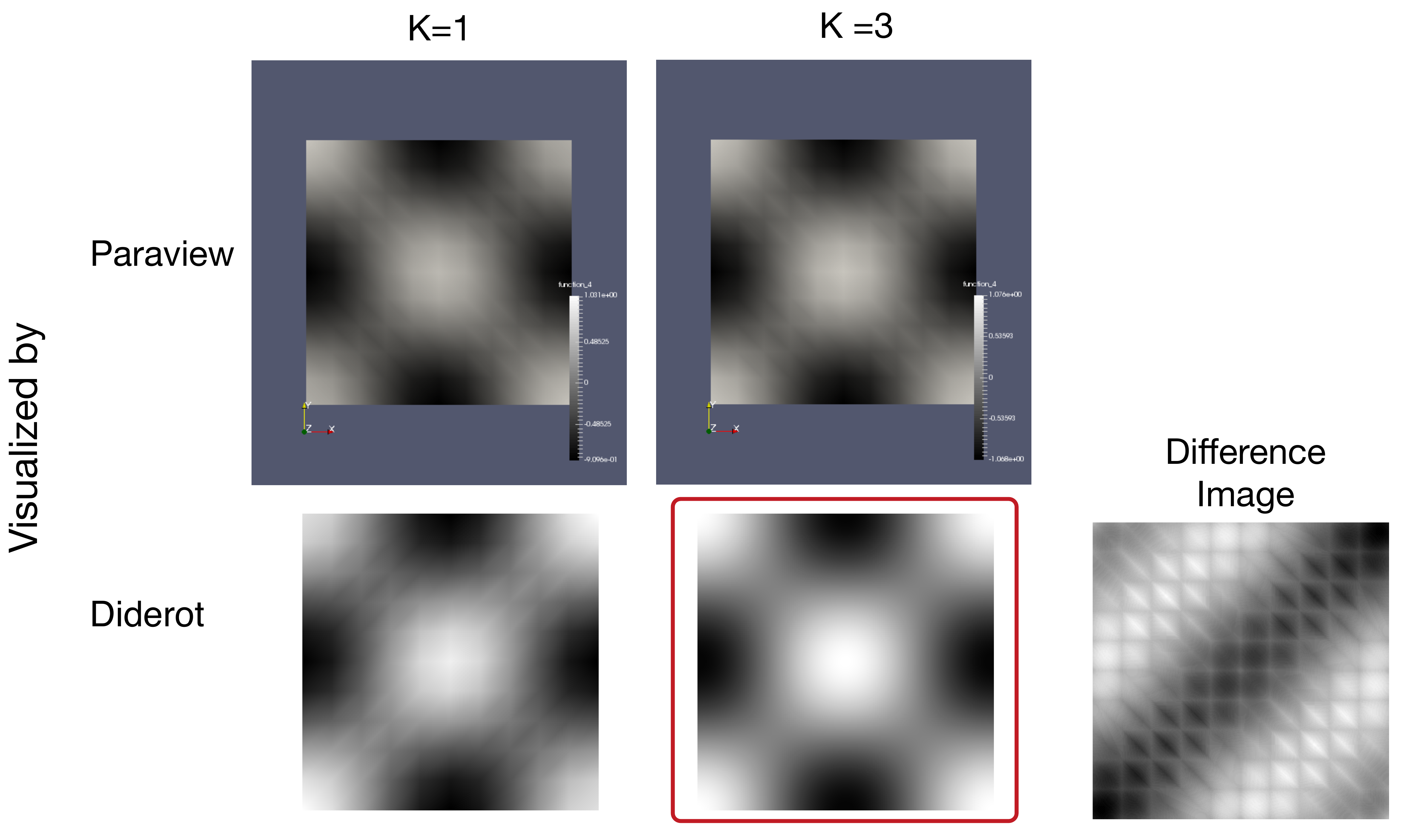} 
\caption[Visualize PDE with Diderot] {The data for the Helmholtz problem is created by Firedrake and visualized by Paraview (top) and Diderot (bottom).}
\label{pde:Helmholtz}
\end{figure}

We illustrate the results using Paraview and Diderot.
The results  are shown in \figref{pde:Helmholtz}.
The image on the bottom right is a difference image and illustrates the difference between the Diderot results.
The data with higher-order elements and visualized with Diderot  is the more clear and refined.
 There is a smoothness captured with the higher-order data that is not in the linear data.

\section{Conclusion}
\label{fem:discuss}
\label{fem:limit}
The Finite-element community has difficulty visualizing fields that use higher order elements.
The communication pipeline allowed us  to demonstrate Diderot's ability to visualize FEM  data by sampling the data and doing volume renderings.
We have shown that Diderot can be used to correctly visualize fields created with higher order elements and a small number of cells.
We have also shown a way to evaluate the results for simple test cases by comparing the result with an external data representation.

It would be beneficial if a Firedrake user could use Diderot with little hassle.
We proposed that we could ease this transition  by providing a Diderot template (or existing Diderot programs) that can be initialized by Firedrake code,
however, as we have shown in the example in \secref{fem:comparediderot} this approach can still lead to errors. 
It takes care, by  the user to ensure that a Diderot program is set-up to correctly visualize the Firedrake data.

We have successfully established communication between Firedrake and Diderot but 
it is limited. 
We suggest further ideas for future work that help develop Diderot and benefit the community.

\paragraph{Debugging FE data with Diderot}
The work on visualizing FE data opens the door to more useful applications. 
  Diderot could potentially be used to visually debug and validate fields created by FEM and possibly reveal hidden details in data created with higher-order elements. 
  We have shown an example of using Diderot to correctly visualize a field created by Firedrake, but have not explored its full potential.

\paragraph{Mature visualization for FE data}
We have not visualized FE data with Diderot programs that use higher-order code.
Partly, because the Diderot compiler cannot yet represent FE data in more intimate level. 
In the future, we look forward to being able to apply more complicated visualization programs to FE fields.

The callback that Firedrake offers, to evaluate a field at a point, is also limited.  
As far as we know, Firedrake does not offer a call back to take third or fourth derivatives of fields (which is necessary in some visualization programs as shown in [13].
Additionally, the call back can only evaluate some fields.
There is no callback supported for some niche problems such as extruded meshes, manifolds, and mixed element meshes.
These type of structures could possibly require more syntax consideration on the Diderot side as well.

\paragraph{Better communication between Diderot and FEM }
A call is made to Firedrake every time a field needs to be evaluated, which makes the communication process expensive.
Each function call creates multiple tensor operations in order to do the right transformations and find the right cell.
The change in coordinates from a reference element to one being computed involves the calculation of the Jacobian matrix, its determinant, and its inverse.

These operations can be similar to previous calls, leading to  redundant and expensive computations.
Diderot does not know how to evaluate a FE field at a point or have the syntax to find the right cell, but it can represent tensor computations.
If we are able to move some of the computations into Diderot then  Diderot can catch these redundant computations and the entire process is less expensive. 
%
%
%
\section*{References}
\begin{enumerate}
\item James Ahrens, Berk Geveci, and Charles Law. Paraview: An end-user tool for large data visualization. In Visualization Handbook, pages 717–731. Academic Press, Inc., Orlando, FL, USA, 2005.
\item Martin S. Alnaes, Anders Logg, Kristian B. Oelgaard, Marie E. Rognes, and Garth N. Wells. Unified form language: A domain-specific language for weak formulations of partial differential equations. ACM Transactions on Mathematical Software, 40, February 2014.
\item A.Logg, K. B. Ølgaard, M. E. Rognes, and G. N. Wells. FFC: the fenics form compiler. In A. Logg, K.A. Mardal, and G. N. Wells, editors, Automated Solution of Differential Equations by the Finite Element Method, volume 84 of Lecture Notes in Computational Science and Engineering, chapter 11, pages 227–238. Springer, 2012.
\item
 Susanne C Brenner and Ridgway Scott. The mathematical theory of finite element methods, volume 15. Springer, 2007.
\item Charisee Chiw. Ein notation in diderot. Master’s thesis, University of Chicago, April 2014.
\item Charisee Chiw, Gordon L Kindlman, and John Reppy. EIN: An intermediate representation for compiling tensor calculus. In Proceedings of the 19th Workshop on Compilers for Parallel Computing (CPC 2019), July 2016.
\item Charisee Chiw, Gordon Kindlmann, and John Reppy. Datm: diderot’s automated testing model. In Proceedings of the 12th International Workshop on Automation of Software Testing, pages 45–51. IEEE Press, 2017.
\item  Charisee Chiw, Gordon Kindlmann, John Reppy, Lamont Samuels, and Nick Seltzer. Diderot: A parallel dsl for image analysis and visualization. SIGPLAN Not., 47(6):111–120, June 2012.
\item Charisee Chiw and John Reppy. Properties of normalization for a math based intermediate representation. arXiv preprint arXiv:1705.08801, 2017.
\item Todd Dupont, Johan Hoffman, Claus Johnson, Robert C Kirby, Mats G Larson, Anders Logg, and R Scott. The FEniCS project. Chalmers Finite Element Centre, Chalmers University of Technology, 2003.
\item  Firedrake. Firedrake: Point evaluation. http://firedrakeproject.org/point-evaluation.html.
\item Luis Ibanez and Will Schroeder. The ITK Software Guide. Kitware Inc., 2005.
\item  Gordon Kindlmann, Charisee Chiw, Nicholas Seltzer, Lamont Samuels, and John Reppy. Diderot: a domain-specific language for portable parallel scientific visualization and image analysis. IEEE Transactions on Visualization and Computer Graphics (Proceedings VIS 2015), 22(1):867–876, January 2016.
\item Gordon L. Kindlmann and Carlos Eduardo Scheidegger. An algebraic process for visualization design. IEEE Trans. Vis. Comput. Graph., 20(12):2181–2190, 2014.
\item Robert C Kirby. Algorithm 839: Fiat, a new paradigm for computing finite element basis functions. ACM Transactions on Mathematical Software (TOMS), 30(4):502–516, 2004.
\item Robert C. Kirby, Matthew Knepley, Anders Logg, and L. Ridgway Scott. Optimizing the evaluation of finite element matrices. SIAM J. Sci. Comput., 27(3):741–758, October 2005.
\item Anders Logg, Kent-Andre Mardal, and Garth Wells. Automated Solution of Differential Equations by the Finite Element Method: The FEniCS Book. Springer Publishing Company, Incorporated, 2012.
\item Fabio Luporini, Ana Lucia Varbanescu, Florian Rathgeber, Gheorghe-Teodor Bercea, J. Ramanujam, David A. Ham, and Paul H. J. Kelly. Cross-loop optimization of arithmetic intensity for finite element local assembly. TACO, 11(4):57:1–57:25, 2014.
\item  Florian Rathgeber, David A. Ham, Lawrence Mitchell, Michael Lange, Fabio Luporini, Andrew T. T. McRae, Gheorghe-Teodor Bercea, Graham R. Markall, and Paul H. J. Kelly. Firedrake: automating the finite element method by composing abstractions. Submitted to ACM TOMS, 2015.
\item Florian Rathgeber, Graham R Markall, Lawrence Mitchell, Nicolas Loriant, David A Ham, Carlo Bertolli, and Paul HJ Kelly. Pyop2: A high-level framework for performance portable simulations on unstructured meshes. In High Performance Computing, Networking, Storage and Analysis (SCC), 2012 SC Companion:, pages 1116–1123. IEEE, 2012.
\item W. Schroeder, K. Martin, and B. Lorensen. The Visualization Toolkit: An Object Oriented Approach to 3D Graphics. Kitware, Inc., Clifton Park, New York, 3rd edition, 2004.
\end{enumerate}
\end{document}